\documentclass[12pt]{article}
\usepackage{graphicx}
\usepackage{epsfig,rotate,multicol,graphicx,psfrag,bm,times}
\usepackage[
            body={15true cm,22true cm},
            twosideshift=0 pt,
            ]{geometry}
\begin{document}

\title{Non-leptonic Weak Interaction in Magnetized Quark matter }
\author{Zheng Xiaoping \thanks{email address:
zhxp@phy.ccnu.edu.cn}, Zhou Xia,Liu Xuewen
\\
{\small  Institute of Astrophysics, Huazhong Normal University,
Wuhan430079 P. R. China }}
\date{}
\maketitle
\begin{abstract}
  We investigated the non-leptonic weak interaction in magnetic field.
  We discussed an improvement of previous method to analytical work out
   the rate for weak field case.Our result easily goes over to field-free limit.Then
  we calculated the reaction rate
  in strong magnetic field where the charged particles are confined
  to the lowest Landau level.A strong magnetic field strongly suppressed the rate,which will be
  foreseen to affect viscous dynamics in SQM .We also derived a
  few approximation formulae under given conditions that can be
  conveniently applied.

\end{abstract}

\section{Introduction}

The composition of comparable number of u,d,and s-quarks are known
as strange quark matter(SQM) would be stable or metastable
configuration of hadronic matter. The bulk SQM is the
$\beta$-equilibrium system determined by such a serious of weak
process: $u+d\rightarrow s+u$, $d\rightarrow
u+e+\overline{\nu}_{e}$ ,$s\rightarrow u+e^{-}+\overline{\nu}_{e}$
\cite{1}. There has been a lot of interest in the study of the
reaction rate and their astrophysical relevance\cite{2,3,4,5,6,7}.
In the interior of neutron stars,the semi-leptonic reaction is
devoted to the cooling of stars while the non-leptonic one to the
damping of instability of rapidly rotating stars. We here concern
about the non-leptonic process. The previous work\cite{2,3} were
carried out under the usual situation of zero magnetic field.
However, the strength of the surface magnetic field of a pulsar is
typically of order $10^{12}G$ which concluded from the observation
data. Some magnetars are observed to have magnetic field of
$10^{14}\sim 10^{15}G$ . Considering the flux throughout the stars
as a simple trapped primordial flux, the internal magnetic field
may go up to $10^{18}G$ or even more \cite{4,5}. In spite of the
fact that we do not know yet any appropriate mechanism to produce
more intense field, the scalar viral theorem indeed allows the
field magnitude to be as large as $10^{20}G$. Therefore,it is
advisable to study the effect of the magnetic field on strange
quark matter including the calculation of the rate of the
reactions. The quark Urca process have been discussed in a few of
works\cite{8,9}.Here we focus on the non-leptonic process leading
to bulk viscosity in strange quark matter.
\begin{equation}
u(1)+d \rightarrow u(2)+s
\end{equation}

In magnetic fields,a classical charged particle will have a
transverse cycle motion. In quantum mechanism, the transverse
motion is quantized into Landau levels. The quantization effect is
important when the magnetic strength is equal or larger than the
critical value $B_{m}^{(c)}={{m_{i}^{2}c^3}/{(q_{i}\hbar)}}$
defined by equating the cyclotron energy $qB/(mc)$ to $mc^{2}$ .
where $m_{i}$ and $q_{i}$ denoted the mass and charge(absolute
value)of the particle \cite{10,11}. $\hbar$,$k$ and $c$ denotes
the Planck constant and Boltzman constant and velocity of light
respectively, which are taken to be units below.We have considered
a wide range of magnetic fields in our study, from "low" to "very
high" magnetic fields($B\geq 10^{19}G$)called respectively "weak"
and "strong" field. Since u-quark mass in magnitude of order is
smaller than d-quark and s-quark,we assume quantization effect on
u-quark is important but that of other flavors are
negligible\cite{4}. In the weak field strength situation,we deal
with the calculation of the rate through some approximations:
(i)the unaffected matrix element; (ii)the approximating
free-particle motion direction. In degeneracy,the strong magnetic
fields,i.e. $2qB>p_{F_{i}}^{2}$, force u-quark to occupy the
lowest Landau ground state, where $p_{F_{i}}$ is the Fermi
momentum,we should use the exact solution of Dirac equation in
magnetic field to evaluate the reaction rate.

This paper is organized as follows. We solve the the dirac
equation in part2. We consider the case of weak field in section
3. And then give the reaction rate in strong field in section 4.
Finally,the results are discussed and a summery is made in section
5.

\section{The solution of Dirac function in magnetic field}
We first solve Dirac equation in the presence of a magnetic field
$\textbf{B}$.Let the uniform magnetic field $\textbf{B}$ along the
z-axis,and we choose the asymmetric $\emph{Landau gauge}$
\begin{equation}
\mathbf{A}=(0, Bx, 0)
\end{equation}
so that the four-dimensional polarized wave function can be
expressed in term of stationary states in the normalization volume
$V=L_{x}L_{y}L_{z}$. In the magnetic field, the u-quark wave
function for fermions ultra-relativistically reads
\begin{equation}
\psi_{+}(t,\mathbf{x})=\frac{exp[-i  E t + i p_y y + i
 p_z z]}{\sqrt{2E(E+m)L_y
L_z}}\left(
\begin{array}{ccc}(E+m)I_{\nu ;p_z}(x)\\
0\\
 p_z I_{\nu ;p_z}(x)\\
-i \sqrt{2\nu qB} I_{\nu -1;p_z}(x) \end{array} \right)
\end{equation}
for spin up, and
\begin{equation}
\psi_{-}(t,\mathbf{x})=\frac{exp[-i E t + i p_y + i
 p_z]}{\sqrt{2E(E+m)L_y
L_z}}\left(
\begin{array}{ccc}
0\\
(E+m)I_{\nu -1;p_z}(x)\\

i \sqrt{2\nu qB} I_{\nu ;p_z}(x)\\
- p_z I_{\nu -1;p_z}(x)\\
\end{array} \right)
\end{equation}
for spin down cases. where the energy $E=\sqrt{p^2_z+m^2+2\nu qB}$
, and $\nu$ is the Landau level.

It is degenerate and can be expressed as other quantum numbers
such as, $l$ is the orbital quantum number and $s$ is the spin
quantum number.
\begin{equation}
\nu=l+\frac{1}{2}(1 -s)
\end{equation}
$q$ is the charge of u-quark. In $Eqn(3),(4)$,
\begin{equation}
\xi=\sqrt{qB}(-x+\frac{p_{y}}{qB})
\end{equation}
and
\begin{equation}
I_{\nu ;p_z}=\left( \frac{qB}{\pi}
\right)^{\frac{1}{4}}exp(-\frac{\xi^2}{2})\times
\frac{1}{\sqrt{2^\nu  \nu !}}H_\nu (\xi)
\end{equation}
where $H_\nu$ is Hermite polynomial.

\section{The rate in weak magnetic field}
If the matrix element for the reaction(1) is determined, we can
express the rate per volume of reaction(1) as\cite{2,3}:
\begin{equation}
\Gamma(u_{1}d\rightarrow
su_{2})=\frac{36}{2}[\prod_{i}\int\frac{d^{3}p_{i}}{(2\pi)^{3}2E_{i}}]|M_{s}|^{2}S(2\pi)^{4}\delta^{4}(P_{1}+P_{d}-P_{2}-P_{s})
\end{equation}
where the phase space integrals are to be calculated over all
particle states, the statistical distribution function
$S=f_{1}f_{d}(1-f_{2})(1-f_{s})$, the quark in equation(1) are
described by Fermi-Dirac distributions in the form
\begin{equation}
f_{i}(E_{i})=[1+\exp(\frac{E_{i}-\mu_{i}}{T})]^{-1}, i=1,2,d,s
\end{equation}
where $\mu_{i}$ are the chemical potential.

We consider  the reaction $(1)$ when the magnetic field is not
strong enough to force the quark into the lowest Landau level.
Previous study \cite{5} shows that the matrix element for the weak
process remains unaffected and only the phase factor is modified.

The matrix element summed over final spins and averaged over
initial spins is given by \cite{2}:
\begin{equation}
|M_{s}|^{2}=64G_{F}^{2}\sin^{2}\theta_{c}\cos^{2}\theta_{c}(P_{1}\cdot
P_{d})(P_{2}\cdot P_{s})
\end{equation}
here $P_{i}=(E_{i}-\textbf{p}_{i})$ is the four-momentum of the
quark $i$ and $G_{F}=1.166 \times 10^{-11}Mev^{-2}$ is the Fermi
constant , $\theta_{c}$ is the Cabibbo
angle($\cos^{2}\theta_{c}=0.948$).We neglect the masses of up and
down quarks, then $E_{1}=p_{1} ,E_{2}=p_{2} ,E_{d}=p_{d},
E_{s}=(p_{s}^{2}+m_{s}^{2})^{1/2}$. so the four-momentum products
can be written as\cite{2,3}
\begin{equation}
(P_{1}\cdot P_{d})(P_{2}\cdot
P_{s})=E_{1}E_{2}E_{d}E_{s}(1-\cos\theta_{1d})(1-\frac{p_{s}}{E_{s}}\cos\theta_{2s})
\end{equation}
where $\theta_{ij}$ denotes the angle between quarks $i$ and $j$ .

Now we can calculate the rate of weak process in the weak magnetic
field by replacing the u-quark phase space factor\cite{5}
\begin{equation}
2\int\frac{d^{3}\textbf{p}}{(2\pi)^{3}}\longrightarrow\frac{qB}{(2\pi)^{2}}\sum^{\nu_{max}}_{\nu=0}(2-\delta_{\nu
,0})\int dp_{z}
\end{equation}
so the reaction rate can be write as:
\begin{eqnarray}
\Gamma(u_{1}d\rightarrow
su_{2})=\frac{18}{(2\pi)^6}G_{F}^{2}\sin^{2}\theta_{c}\cos^{2}\theta_{c}(eB)^{2}\sum_{\nu_{1}=0}^{\nu_{max}}(2-\delta_{\nu_{1,0}})\sum_{\nu_{2}=0}^{\nu_{max}}(2-\delta_{\nu_{2,0}})\nonumber
\\\int p_{d}^{2}dp_{d}p_{s}^{2}dp_{s}\int dp_{1z}\int
dp_{2z} S \delta(E_{1}+E_{d}-E_{2}-E_{s})I
\end{eqnarray}
where
\begin{equation}
I=\int(\prod^{d,s}_{i})d\Omega_{i}(1-\cos\theta_{1d})(1-\frac{p_{s}}{E_{s}}\cos\theta_{2s})\delta^{3}(\textbf{p}_{1}+\textbf{p}_{d}-\textbf{p}_{2}-\textbf{p}_{s})
\end{equation}

As we known ,the angle between two vector is a function of the
respective inclinations and azimuth angle in spherical
coordinates, ie.
\begin{eqnarray}
\cos\theta_{1d}=\cos\theta_{1}\cos\theta_{d}+\sin\theta_{1}\sin\theta_{d}\cos(\varphi_{1}-\varphi_{d})
\nonumber
\\\cos\theta_{2s}=\cos\theta_{2}\cos\theta_{s}+\sin\theta_{2}\sin\theta_{s}\cos(\varphi_{2}-\varphi_{s})
\end{eqnarray}
In magnetic field,the angles of the polarized
quarks,$\theta_{1},\theta_{2},\varphi_{1},\varphi_{2}$ vary with
Landau level ($\nu_{1},\nu_{2}$) and z-component momentum
($p_{1z},p_{2z}$) . Therefore the integrals and summations in
$Eqn(12)$ become very difficult due to the coupling of variables
$\nu$ and $p_{z}$ into the integrated function. We need to make a
improvement on chakrabaty's approach. Fortunately, we can
approximately regard the angles as independent variables for the
weak field situation and then the integrals and summations
decouples each other, because the kinetic quark direction should
have only a small deviation from the field-free case,although the
modification of the absolute value of the momentum considered
under the situations.It completely coincides with free-particle's
matrix approximation described by $Eqn(10)$.

We immediately have:
\begin{equation}
2\int\frac{d^{3}\textbf{p}}{(2\pi)^{3}}\longrightarrow\frac{qB}{(2\pi)^{3}}\sum\limits^{\nu_{max}^{'}}_{\nu=0}(2-\delta_{\nu
,0})\int' dp_{z}\int' d\Omega
\end{equation}
where $\sum'$ and $\int'$ denote the summation and integral
independent of the direction angles. In this approximation , which
may be called free-particle direction average(FDA). Eqn(13) and
Eqn(14) read:
\begin{eqnarray}
\Gamma(u_{1}d\rightarrow
su_{2})=\frac{18}{(2\pi)^8}G_{F}^{2}\sin^{2}\theta_{c}\cos^{2}\theta_{c}(eB)^{2}\sum_{\nu_{1}=0}^{\nu_{1max}}(2-\delta_{\nu_{1,0}})\sum_{\nu_{2}=0}^{\nu_{2max}}(2-\delta_{\nu'_{2,0}})\nonumber
\\\int p_{d}^{2}dp_{d}p_{s}^{2}dp_{s}\int dp_{1z}\int
dp_{2z} S \delta(E_{1}+E_{d}-E_{2}-E_{s})I'
\end{eqnarray}
where
\begin{equation}
I'=\int'(\prod^{1,2,d,s}_{i})d\Omega_{i}(1-\cos\theta_{1d})(1-\frac{p_{s}}{E_{s}}\cos\theta_{2s})\delta^{3}(\textbf{p}_{1}+\textbf{p}_{d}-\textbf{p}_{2}-\textbf{p}_{s})
\end{equation}

 Since $\mu_{i}>>T$ ,only
those fermions whose momenta lie close to their respect Fermi
surface can take part in the reaction. We use the method in
\cite{6,7}to complete partially the integral of $I'$ through
\begin{equation}
\delta^{3}(\textbf{p}_{1}+\textbf{p}_{d}-\textbf{p}_{2}-\textbf{p}_{s})=\int\frac{d^{3}\textbf{x}}{(2\pi^{3})}\exp(i\textbf{p}\cdot\textbf{x})
\end{equation}
can be written as:
\begin{eqnarray}
I' &=&
\frac{2^{7}\pi^{2}}{p_{F_{1}}p_{F_{2}}p_{F_{d}}p_{F_{s}}}\int\frac{dx}{x^{2}}[\prod_{i}^{1,2,d,s}\sin(p_{F_{i}}x)+a\sin(p_{F_{1}}x)\sin(p_{F_{d}}x)f(p_{F_{2}}x)f(p_{F_{s}}x)+\nonumber
\\&&{}\sin(p_{F_{2}}x)\sin(p_{F_{s}}x)f(p_{F_{1}}x)f(p_{F_{d}}x)+a\prod_{i}^{1,2,d,s}f(p_{F_{i}}x))]\
\\&\equiv&\frac{2^{7}\pi^{2}}{p_{F_{1}}p_{F_{2}}p_{F_{s}}}J
\end{eqnarray}
where $a=\frac{p_{F_{s}}}{\mu_{s}} ,
f(p_{F_{i}}x)=\cos(p_{F_{i}}x)-\frac{\sin(p_{F_{i}}x)}{p_{F_{i}x}}$
and the integral $J$ is defined through $Eqn(20)$ and
$Eqn(21)$,can be calculated numerically.

The net rate of transforming d-quark into s-quark is \cite{2}
\begin{equation}
\Gamma(d\rightarrow
s)=[1-\exp(\frac{\mu_{d}-\mu_{s}}{T})]\Gamma(u_{1}d\rightarrow
su_{2})
\end{equation}
Using the method in \cite{7,12} and substitute $Eqn(17),(18)$ and
(21) into (22) we can get
\begin{eqnarray}
\Gamma(d\rightarrow
s)=\frac{3}{2\pi^{6}}G_{F}^{2}\sin^{2}\theta_{c}\cos^{2}\theta_{c}(qB)^{2}
\sum^{\nu_{1max}}_{\nu1=0}(2-\delta_{\nu1
,0})\sum^{\nu_{2max}}_{\nu=0}(2-\delta_{\nu2 ,0})\nonumber
\\
\frac{\mu_{d}^{2}\mu_{s}}{\sqrt{\mu_{1}^{2}-2\nu_{1}qB}\sqrt{\mu_{2}^{2}-2\nu_{2}qB}}\Delta\mu
(\Delta\mu^{2}+4\pi^{2}T^{2})J
\end{eqnarray}
where
\begin{equation}
\nu_{imax}=Int(\frac{\mu_{i}^{2}}{2qB}),i=1,2,
\Delta\mu=\mu_{d}-\mu_{s}
\end{equation}
when the $B\rightarrow 0$ the sum can be replaced by integral of
$\nu$ and then the result goes over to the expression:
\begin{equation}
\Gamma(d\rightarrow
s)=\frac{6}{\pi^{6}}G_{F}^{2}\sin^{2}\theta_{c}\cos^{2}\theta_{c}\mu_{d}^{2}\mu_{u}^{2}\mu_{s}
\Delta\mu (\Delta\mu^{2}+4\pi^{2}T^{2})J
\end{equation}
It is just the field-free case.

\section{The rate of weak process in strong magnetic field}
Now we consider strong magnetic field effect on the non-leptonic
weak interaction in this section.In the case of strong magnetic
field that $B\geq B_{m}^{(u)}$, all u-quarks occupy the lowest
Landau ground state with the u-quark spins pointing in the
direction of the magnetic field.We treat other flavors in this
process as free particle which do not affected by the magnetic
field. The matrix element for the reaction reads:
\begin{equation}
M=\frac{G_{F}\sin\theta_{c}\cos\theta_{c}}{\sqrt{2}}\int\bar{\psi_{2}}\gamma_{\mu}(1-\gamma_{5})\psi_{d}\bar{\psi_{s}}\gamma^{\mu}(1-\gamma_{5})\psi_{1}d^{4}x
\end{equation}
d,s quark treat as free particle ,then we get
\begin{equation}
\psi_{i}=\frac{1}{V^{\frac{1}{2}}}exp(-\emph{i}P_{i}\cdot r)U_{i}
\end{equation}
\begin{equation}
U_{i}=\sqrt{\frac{E_{i}+m_{i}}{2E_{i}}}\left(
\begin{array}{c}1\\
0\\
\frac{ p_{z_{i}}}{E_{i}+m_{i}}\\
\frac{ p_{x_{i}}+ip_{y_{i}}}{E_{i}+m_{i}}
\end{array} \right)
\end{equation}
where $P_{i}$ denotes four-dimensional momentum,$i=d,s$.

Consider $\nu =0$,the wave function $Eqn(3)(4)$ of u-quark can be
writen as
\begin{equation}
\psi_{1}=\frac{1}{\sqrt{L_{y}L_{z}}}exp(-iE_{1}t+ip_{y_{1}}y_{1}+ip_{z_{1}}z_{1})(\frac{qB}{\pi})^{\frac{1}{4}}exp[-\frac{qB}{2}(-x+\frac{p_{y_{1}}}{qB})^{2}]U_{1}
\end{equation}
\begin{equation}
\psi_{2}=\frac{1}{\sqrt{L_{y}L_{z}}}exp(-iE_{2}t+ip_{y_{2}}y_{2}+ip_{z_{2}}z_{2})(\frac{qB}{\pi})^{\frac{1}{4}}exp[-\frac{qB}{2}(-x+\frac{p_{y_{2}}}{qB})^{2}]U_{2}
\end{equation}
where
\begin{equation}
U_{1}=\frac{1}{\sqrt{2E_{1}(E_{1}+m_{u_{1}})}}\left(
\begin{array}{c}E_{1}+m\\
0\\
 p_{z_{1}}\\
0 \end{array} \right)
\end{equation}
\begin{equation}
U_{2}=\frac{1}{\sqrt{2E_{2}(E_{2}+m_{u_{2}})}}\left(
\begin{array}{c}E_{2}+m\\
0\\
 p_{z_{2}}\\
0 \end{array} \right)
\end{equation}
for spin up.

 We now calculate the matrix elements squared and
summed over the initial state and average over the final state.We
use the $Eqn(29),Eqn(32)$ and $Eqn(33)$ to express corresponding
matrix element:

$$|M_{s}|^{2}=[\bar{U_{2}}\gamma_{\mu}(1-\gamma_{5})U_{d}\bar{U_{s}}
\gamma^{\mu}(1-\gamma_{5})U_{1}][\bar{U_{2}}\gamma_{\mu}(1-\gamma_{5})U_{d}\bar{U_{s}}
\gamma^{\mu}(1-\gamma_{5})U_{1}]^{\dag}$$

And then imediately get:
\begin{eqnarray}
|M_{s}|^{2}=\frac{1}{2E_{1}^{2}E_{2}^{2}E_{d}E_{s}}(E_{1}-p_{z_{1}})^{2}(E_{2}+p_{z_{2}})^{2}(E_{d}+p_{z_{d}})(E_{s}-p_{z_{s}})
\end{eqnarray}
Here we set $m_{d}=m_{u}=0$ ,and then
$p_{z_{1}}<0$,$p_{z_{2}}>0$,otherwise $|M_{s}|^{2}=0$. Carrying
out the integral ,we obtain

\begin{eqnarray}
|M|^{2}=\frac{G_{F}^{2}\sin\theta_{c}^{2}\cos\theta_{c}^{2}}{2VL_{x}^{2}(L_{y}L_{y})^{3}}(2\pi)^{3}
|M_{s}|^{2}\exp[\frac{-(p_{y_{1}}-p_{y_{2}})^{2}{-(p_{x_{d}}-p_{x_{s}})^{2}}}{2qB}]\nonumber
\\\delta(E_{2}+E_{s}-E_{1}-E_{d})\delta(p_{y_{1}}+p_{y_{d}}-p_{y_{2}}-p_{y_{s}})
\delta(p_{z_{1}}+p_{z_{d}}-p_{z_{2}}-p_{z_{s}})
\end{eqnarray}

The rate per volume of the reaction is given by:

\begin{eqnarray}
\Gamma(u_{1}d\rightarrow
su_{2})=\frac{n_{1}n_{d}}{2}\int\frac{Vd^{3}\textbf{p}_{d}}{(2\pi)^{3}}\int\frac{Vd^{3}\textbf{p}_{s}}{(2\pi)^{3}}
\int_{\frac{-qBL_{x}}{2}}^{\frac{qBL_{x}}{2}}\frac{L_{y}}{2\pi}dp_{y_{1}}
\int_{\frac{-qBL_{x}}{2}}^{\frac{qBL_{x}}{2}}\frac{L_{y}}{2\pi}dp_{y_{2}}\nonumber
\\
\int_{-\infty}^{\infty}\frac{L_{y}}{2\pi}dp_{z_{1}}\int_{-\infty}^{\infty}\frac{L_{y}}{2\pi}dp_{z_{2}}
|M|^{2}f_{1}f_{d}(1-f_{2})(1-f_{s})
\end{eqnarray}
where the factor $n_{1}=n_{d}=6$ comes from 2 spins and 3
colors.Since only left-hand helicity states of $u_{1}-quark$
couples to the $W^{-}$ ($W^{-}$ is the mediate of the reaction), a
factor of $\frac{1}{2}$ is shown in the $Eqn(35)$.  The integrals
over $dp_{y_{1}}$ and $dp_{y_{2}}$ can be carried out using the
y-component momentum delta function. The integrals over
$dp_{z_{1}}$ and $dp_{z_{2}}$ are converted into $dE_{1}$ and
$dE_{2}$ respectively\cite{4,5}.

Since $\mu_{s}>>T$,only those momenta lie close to their
respective Fermi surfaces can take part in the reaction. As the
z-component momentum conservation,we can get
\begin{equation}
p_{F_{1}}+p_{F_{d}}\cos\theta_{d}-p_{F_{2}}-p_{F_{s}}\cos\theta_{s}=0
\end{equation}
We approximately set $p_{F_{u}}=p_{F_{d}}=p_{F_{s}}$ near the
equilibrium.Then we can get $\cos\theta_{d}-\cos\theta_{s}=2$,so
we can take out the integral over d-quark and s-quark's momentum
space.

Substituting Eqn(35) into Eqn(22), we obtain

\begin{eqnarray}
\Gamma(d\rightarrow
s)=\frac{G_{F}^{2}\sin^{2}\theta_{c}\cos^{2}\theta_{c}(q
B)}{4\pi^{5}}\exp[\frac{(2\mu_{u})^{2}-(\mu_{d}+p_{F_{s}})^{2}}{2qB}]\nonumber
\\(3\mu_{s}+6\mu_{u}-\mu_{d}) \mu_{d}^{2}\Delta\mu (\Delta\mu^{2}+4\pi^{2}T^{2})
\end{eqnarray}
where
$\Delta\mu=\mu_{d}-\mu_{s}$,$p_{F_{s}}=\sqrt{\mu_{s}^{2}-m_{s}^{2}}$.

\section{Discussion and Conclusion}

We have done calculations of the rate of non-leptonic quark weak
process in magnetic field and give the analytic solutions under
weak-field and strong-field approximation. Based on these results
, we express the net rate of the transforming d-quark to s-quark
in a unified form:
\begin{equation}
\Gamma(d\rightarrow s)=\Gamma_{k}(n_{b},qB)\Delta\mu
(\Delta\mu^{2}+4\pi^{2}T^{2})
\end{equation}
where $k$ takes $0,L,H$ denotes zero-field, weak-field and
strong-field cases respectively.In according with the formula(25),
the result go over to the field-free case when the magnetic field
strength vanishes. We thus have:
\begin{equation}
\Gamma_{0}(n_{b})=\frac{16}{5}G_{F}^{2}\sin^{2}\theta_{c}\cos^{2}\theta_{c}(\frac{n_{b}}{\pi})^{\frac{5}{3}}
\end{equation}
meanwhile for $2q_{u}B<\mu_{u}^{2}$,the $\Gamma_{L}$ reads simply
from Eqn(23)
\begin{equation}
\Gamma_{L}(n_{b},qB)=\frac{144}{\pi^{4}}G_{F}^{2}\sin^{2}\theta_{c}\cos^{2}\theta_{c}q_{u}Bn_{b}J\frac{\nu_{m}^{\frac{3}{2}}}{3\nu_{m}^{\frac{1}{2}}+4(\nu_{m}-1)^{\frac{3}{2}}+8\nu_{m}^{\frac{3}{2}}}
\end{equation}
where $J$ as a function of $\mu$ define in figure(1).We find
$\Gamma_{L}$ has a small deviation from field-free case.The
comparison is made in figure(2).

The simplified formula of $\Gamma_{H}(n_{b},qB)$ need slightly
complicated calculations for $2q_{u}B\geq\mu_{u}^{2}$

In the referred SQM ($\mu_{i}\gg m_{i}$,only u-quark
polarized)here , the number density of the components read:
\begin{equation}
n_{d,s}=\frac{\mu_{d,s}^{3}}{\pi^{2}}\nonumber
\end{equation}

\begin{equation}
n_{u}=\frac{3q_{u}B\mu_{u}}{2\pi^{2}}
\end{equation}
the $\beta$-equilibrium
\begin{equation}
\mu_{d}=\mu_{s}=\mu,\mu_{u}=\mu-\mu_{e}
\end{equation}
the charge neutrality
\begin{equation}
2n_{u}-n_{d}-n_{s}-3n_{e}=0
\end{equation}
the baryon number density conservation
\begin{equation}
n_{b}=\frac{1}{3}(n_{d}+n_{u}+n_{s})
\end{equation} should be satisfied.
Then we solve these equation numerically to obtain the chemical
potential of the quarks with respective magnetic field.

Figure(3) shows that the chemical potentials of quarks are nearly
equal under this situation.Combines $Eqn(41)(42)(44)$ and $(45)$
in the approximation of $\mu_{u}\approx\mu_{d}=\mu_{s}=\mu$ ,$\mu$
can be solved analytically through the algebraic equation

\begin{equation}
\mu^{3}+\frac{3}{4}q_{u}B\mu-\frac{\pi^{2}}{2}n_{b}=0
\end{equation}
and then $\Gamma_{H}(n_{b},qB)$ is arrived at
\begin{eqnarray}
\Gamma_{H} &=&
\frac{G_{F}^{2}\sin^{2}\theta_{c}\cos^{2}\theta_{c}(q
B)}{4\pi^{5}}[\frac{q_{u}B}{(-6n_{b}\pi^{2}+\sqrt{36n_{b}^{2}\pi^{4}+(q_{u}B)^{3}})^{\frac{1}{3}}}-\nonumber
\\&&{}(-6n_{b}\pi^{2}+\sqrt{36n_{b}^{2}\pi^{4}+(q_{u}B)^{3}})^{\frac{1}{3}}]^{3}
\end{eqnarray}

Figure(4) give a comparison of $Eqn(47)$ with $Eqn(37)$. They fit
well each other with a small error.

The formula (47) will reduce to
\begin{equation}
\Gamma_{H}=\frac{24G_{F}^{2}\sin^{2}\theta_{c}\cos^{2}\theta_{c}}
{11\pi^{3}}q_{u}Bn_{b }
\end{equation}
for $2qB\sim\mu^{2}$,and tend to the limit
\begin{equation}
\Gamma_{H}=\frac{16\pi
G_{F}^{2}\sin^{2}\theta_{c}\cos^{2}\theta_{c}n_{b}^{3}}{(q_{u}B)^{2}}
\end{equation}
for $2qB\gg\mu^{2}$.

Figure(5) give a comparison of $Eqn(47)$ with $Eqn(48)$ and
$Eqn(49)$. We find they will be some good approximations in future
realistic application.

Under the consideration that u-quark is polarized but effect of
other flavors are negligible, we investigate the influence of
magnetic field on the non-leptonic rate although the result for
the weak field case has a small deviation from the field-free case
,we give an analytical treatment of the weak reaction which can be
extended to the calculations of other reaction process. However,
the strong magnetic field can extremely suppress the rate.It is
possible to lead the decrease of bulk viscosity in magnetized
SQM.This may have serious implications for compact star and pulsar
dynamics.In fact, we should also distinguish d-quark and s-quark
in calculations to obtain refining result for applications. It is
our future works.

This work is supported by NFSC under Grant Nos.90303007 and
10373007,and the Ministry of Education of China with project
No.704035.

\begin{figure}[thp]
\includegraphics[bb=0 10 100 200]{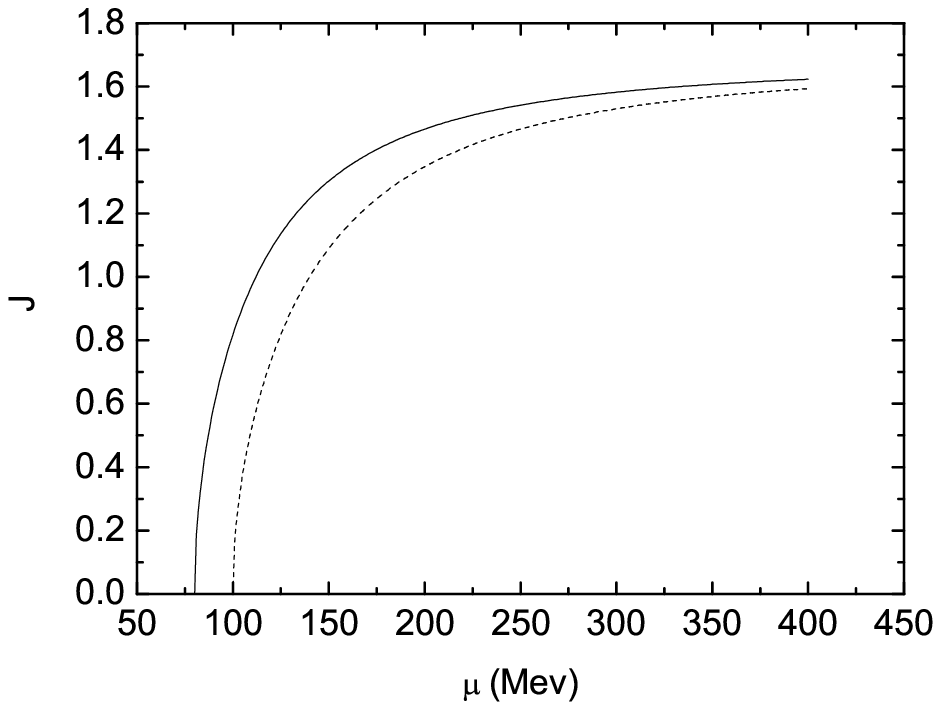}
\caption {numerical result of J , given by $Eqn(20)$ and $(21)$,
show as as a function of $\mu$ for various value of the parameters
$m_{s}$. The solid curve is for $m_{s}=80Mev$,The dot curve is for
$m_{s}=100Mev$.}\label{fig1}
\end{figure}
\begin{figure}[hp]
\includegraphics[bb=0 10 100 200]{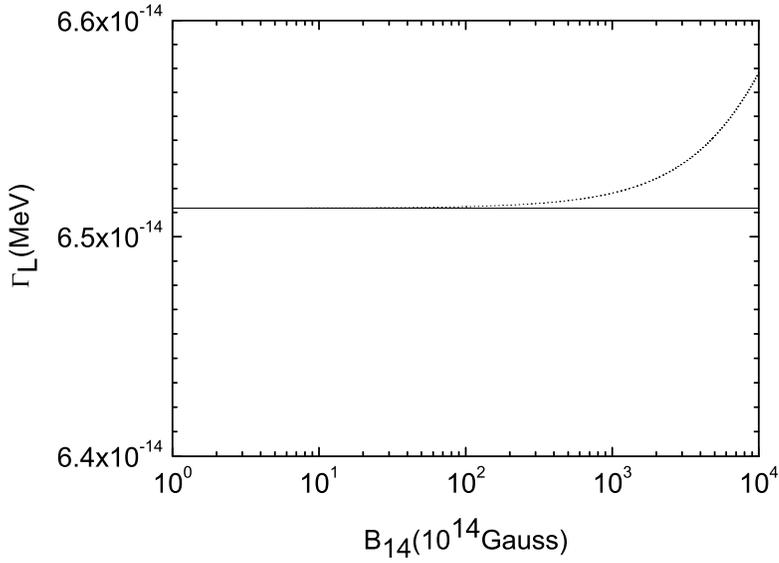}
\caption {Derivation of $\Gamma_{L}$ from $\Gamma_{0}$ with $B$
when $n_{b}=0.2fm^{-3}$.}\label{fig2}
\end{figure}
\begin{figure}[bp]
\includegraphics[bb=0 10 100 200]{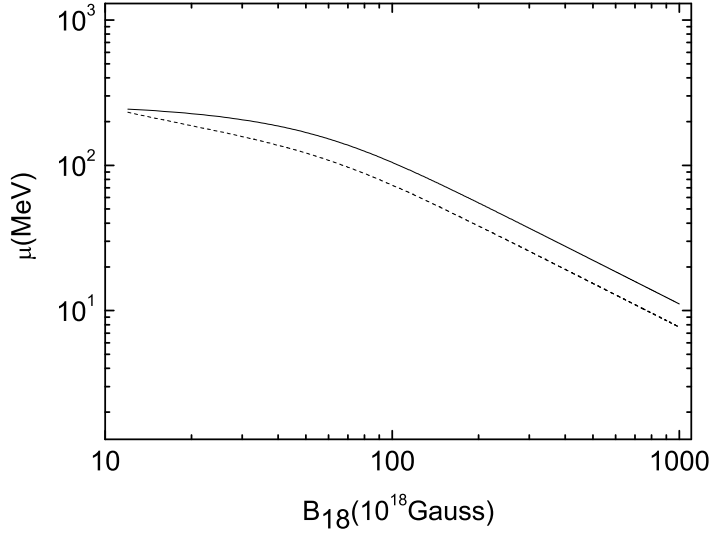}
\caption {numerical result of $\mu$ as a function of $\textbf{B}$
which get from $Eqn(46)(47)(48)$.The dot curve is for $\mu_{u}$
when $n_{B}=0.2fm^{-3}$.The solid curve is for the case when
$\mu=\mu_{d}=\mu_{s}$}.\label{fig3}
\end{figure}

\begin{figure}[hp]
\includegraphics[bb=0 10 100 200]{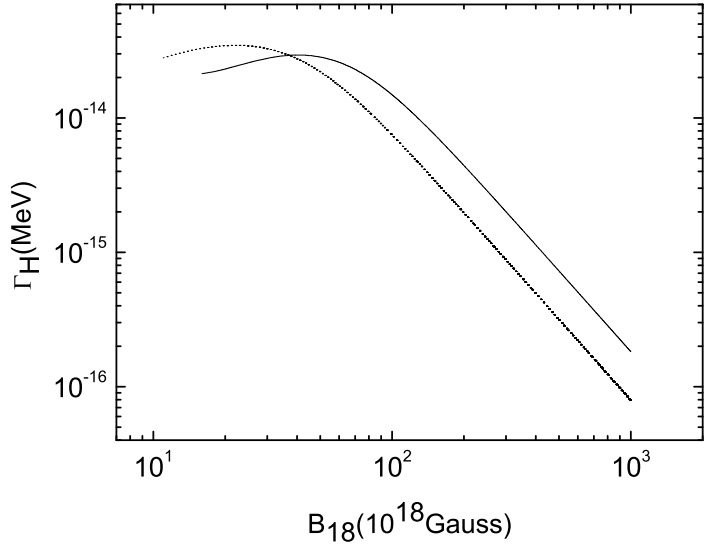}
\caption {$\Gamma_{H}$ as a function of $B$ when
$n_{b}=0.2fm^{-3}$.The solid curve is for the result of  $Eqn(37)$
and the dot curve is for the result of $Eqn(47)$.}\label{fig4}
\end{figure}
\begin{figure}[hp]
\includegraphics[bb=0 10 100 200]{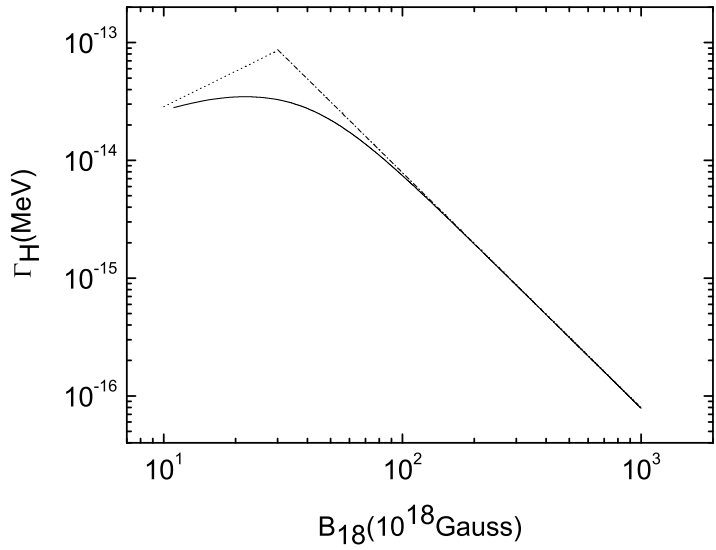}
\caption {$\Gamma_{H}$ as a function of $B$ when
$n_{b}=0.2fm^{-3}$.The solid curve is for the result of  $Eqn(47)$
and the dot curve is for the result of $Eqn(48)$,the dash dot
curve is for the result of $Eqn(49)$.}\label{fig5}
\end{figure}

\end{document}